\documentclass[pra,twocolumn,showpacs,amsmath,amssymb,superscriptaddress]{revtex4}
\usepackage{graphicx}
\usepackage{bm}
\usepackage{amsmath,amssymb}
\begin{document}
\title{Entanglement-fidelity relations for inaccurate ancilla-driven quantum computation}
\author{Tomoyuki Morimae}
\email{morimae@gmail.com}
\affiliation{
Laboratoire Paul Painlev\'e, Universit\'e Lille 1,
F-59655 Villeneuve d'Ascq Cedex, France}
\author{Jonas Kahn}
\affiliation{
Laboratoire Paul Painlev\'e, Universit\'e Lille 1,
F-59655 Villeneuve d'Ascq Cedex, France}
\affiliation{Centre National de la Recherche Scientifique, France}
\date{\today}
\begin{abstract}
It was shown in [T. Morimae, Phys. Rev. A {\bf81}, 060307(R) (2010)]
that the gate fidelity of an inaccurate one-way quantum computation
is upper bounded by a decreasing function of 
the amount of entanglement in the register. This means that a 
strong entanglement causes the low gate fidelity in the one-way 
quantum computation with inaccurate measurements.
In this paper, 
we derive similar entanglement-fidelity relations  
for the inaccurate ancilla-driven quantum computation.
These relations again imply that a strong entanglement in the register 
causes the low gate fidelity in the ancilla-driven quantum computation
if the measurements on the ancilla are inaccurate.
\end{abstract}
\maketitle  
\section{Introduction}
\label{introduction}
In the circuit model~\cite{Nielsen} of quantum computation, 
the quantum register which stores quantum information consists of many
qubits, and a quantum gate operation is performed by
directly accessing one or two qubits in the quantum register. 
The most challenging task
for an experimentalist who adopts the circuit model 
is therefore the coherent establishment of entanglement among 
register qubits in parallel with the execution of a quantum algorithm.
From the theoretical point of view, the role of entanglement played
in the circuit model of quantum computation has been the 
most fundamental subject of study~\cite{Jozsa,Vidal}.

In the one-way model~\cite{cluster} of quantum computation,
on the other hand, a highly
entangled state which is called the ``cluster state" (or the ``graph state") is prepared in advance and the whole quantum computation is performed
by adaptive measurements of each qubit.
The preparation of the resource (i.e., entanglement) is thus
clearly separated from
the consumption of the resource.
This great advantage of the one-way model
has lead to many experimental implementations~\cite{e1,e2,e3,e4,e5,e6,e7}
and theoretical investigations about the roles of
entanglement in the one-way quantum 
computation~\cite{Nest,Gross,Bremner,novelcluster,error_cluster}.

Recently, 
a mixture of those two models,
which is called 
the ancilla-driven quantum computation,
was proposed~\cite{ADQC1,ADQC2}.
In this model,
the quantum register is a set of many qubits  
like the circuit model.
However, 
a quantum gate operation
is, like the one-way model, performed by
using entanglement and measurements:
one or two register qubits are coupled to a single mobile ancilla, and the
ancilla is measured after establishing the interaction between the
ancilla and register qubit(s). 
The backaction of this measurement provides
the desired gate operation, such as a single qubit rotation or an 
entangling two-qubit operation, on register qubit(s) 
(see Fig.~\ref{adqc}). 
The main feature of this model is that  
the universal quantum computation is performed with
only a single type of interaction (CZ or CZ+SWAP) between 
the ancilla and register
qubit(s).
It is advantageous to some experimental setups
where the implementation of various types of interactions 
at the same time is very difficult (such as the solid-based quantum
computation) or
where the flying ancilla mediates interactions between static qubits
(such as the chip-based quantum computation~\cite{chip} or the hybrid quantum computation
of matter and optical elements~\cite{hybrid}).

In this paper, we study how entanglement among register qubits
affects the gate fidelity in the ancilla-driven quantum computation
if the measurement is inaccurate.
For this purpose, we generalize the result of
Ref.~\cite{error_cluster} to an inaccurate ancilla-driven quantum computation.
In Ref.~\cite{error_cluster},
the relation 
\begin{eqnarray}
F\le 1-S\sin^2\frac{\epsilon}{2}
\label{similar}
\end{eqnarray}
between entanglement $S$, the gate fidelity $F$,
and the inaccuracy $\epsilon$ of the measurement 
was derived for the inaccurate one-way model 
(for details, see Sec.~\ref{one-way}).
The meaning of this inequality is that if the entanglement is strong,
the inaccurate measurements make the gate fidelity low.

The results of this paper are:
(I) For the ancilla-driven single-qubit rotation, 
we obtain the same entanglement-fidelity relation as given in 
Eq.~(\ref{similar}).
(II) For the ancilla-driven two-qubit entangling gate with the CZ interaction,
we again obtain the same entanglement-fidelity relation as given in
Eq.~(\ref{similar}). 
(III) For the ancilla-driven two-qubit entangling gate with the CZ+SWAP interaction,
we obtain the entanglement-fidelity relation Eq.~(\ref{main2}) which 
is slightly different from Eq.~(\ref{similar}).
However Eq.~(\ref{main2}) also implies that if the entanglement is strong,
the inaccurate measurements make the gate fidelity low.

This paper is organized as follows:
We will briefly review the ancilla-driven quantum
computation~\cite{ADQC1,ADQC2} in Sec.~\ref{ancilla-driven} for the convenience
of the reader.
In Sec.~\ref{one-way},
we will review and extend 
the result of Ref.~\cite{error_cluster}
about the entanglement-fidelity relation for the inaccurate one-way model.
We will study the entanglement-fidelity relation for
the inaccurate
ancilla-driven single-qubit rotation in Sec.~\ref{ancilla-drivenrotation}.
We will also study the entanglement-fidelity relation for the inaccurate
ancilla-driven
two-qubit entangling gates with the CZ interaction 
in Sec.~\ref{ancilla-drivenCZ} 
and that with the CZ+SWAP interaction
in Sec.~\ref{ancilla-drivenCZSWAP}, 
respectively.

Throughout this paper, $\hat{X}_i$, $\hat{Y}_i$, and $\hat{Z}_i$
are Pauli $x$, $y$, and $z$ operators on $i$th qubit, respectively. 
$\hat{1}_i$ is the identity operator on $i$th qubit.
Eigenvectors of them are $\hat{X}|\pm\rangle=\pm|\pm\rangle$
and $\hat{Z}|z\rangle=(-1)^z|z\rangle$ $z\in\{0,1\}$,
respectively. $\hat{H}_i$ is the Hadamard operator acting on $i$th
qubit. The Hadamard operator works as 
$\hat{H}|0\rangle=|+\rangle$
and
$\hat{H}|1\rangle=|-\rangle$.

\section{Ancilla-driven quantum computation}
\label{ancilla-driven}
\begin{figure}[htbp]
\begin{center}
\includegraphics[width=0.5\textwidth]{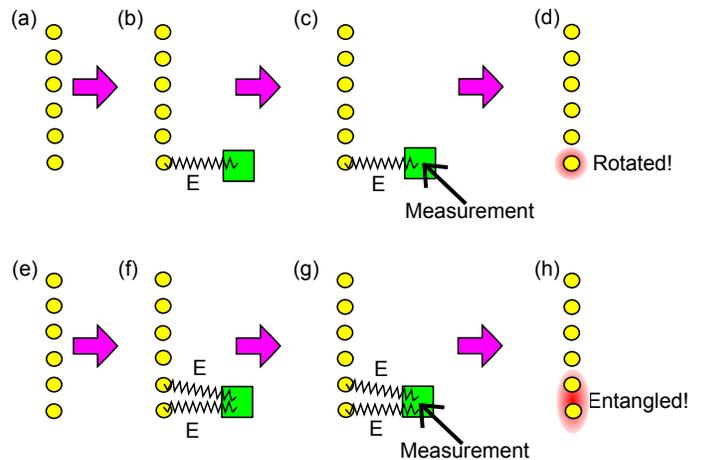}
\end{center}
\caption{(Color online.) The ancilla-driven quantum gates~\cite{ADQC1,ADQC2}.
Yellow circles are register qubits. 
Top line: a single-qubit rotation. (b) The ancilla (green square)
is coupled to the qubit we want to rotate (say, the bottommost one) through the interaction 
$E$, which is represented by the black zigzag line. 
(c) After the interaction, the ancilla
is projected onto a certain direction (represented by the solid black arrow). 
(d) The measurement backaction rotates
the bottommost qubit of the register by the desired angle.
Bottom line: a two-qubit entangling gate. 
(f) The ancilla (green square)
is coupled to two qubits (say, the
two bottommost ones) through the interaction $E$, which is
{\it the same} as the interaction used in the top line. 
The interaction is represented by the black zigzag line.
(g) After the interaction, the ancilla is projected (solid black arrow).
(h) The measurement backaction causes the desired entangling gate between 
the two bottommost qubits of the register.
} 
\label{adqc}
\end{figure}

Let us briefly review the ancilla-driven 
quantum computation~\cite{ADQC1,ADQC2}.
As in the case of the circuit model~\cite{Nielsen},
the quantum register is a set of $N$ qubits.  
Unlike the circuit model, however, a
quantum operation on one or two qubits in the register
is indirectly driven by a single 
mobile ancilla which can couple to one or two qubits
through a fixed interaction (see Fig.~\ref{adqc}). 
An advantage of this model is that  
a single type of interaction is sufficient for
the universal quantum computation. 

\begin{figure}[htbp]
\begin{center}
\includegraphics[width=0.4\textwidth]{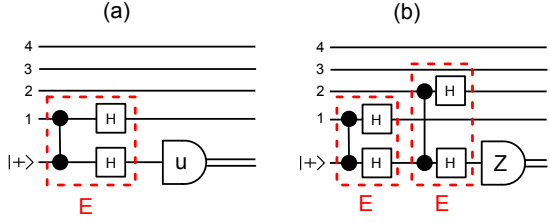}
\end{center}
\caption{(Color online.) Ancilla-driven quantum gates with the CZ interaction~\cite{ADQC1,ADQC2}.
(a) The quantum circuit realizing
the single-qubit rotation $\hat{J}(u)=\hat{H}e^{i\frac{u}{2}\hat{Z}}$.
First, the CZ gate is applied between the qubit 
we want to rotate (say, first qubit)
of the register $|\psi\rangle$ and the ancilla
which is initialized to be $|+\rangle$.
Second, the Hadamard gate is applied to each of them.
Finally, the ancilla is measured in 
$|u_\pm\rangle\equiv(|0\rangle\pm e^{iu}|1\rangle)/\sqrt{2}$ basis.
The backaction of this measurement changes the register state into
$\hat{X}_1^j\hat{J}_1(u)|\psi\rangle$ depending
on the measurement result $j=0,1$, where
the subscript ``1" indicates that the operator is acting on first qubit.
The interaction $\hat{E}=\hat{H}_1\hat{H}_A\hat{CZ}_{1,A}$ is specified with the red box.
(b) The quantum circuit realizing the CZ gate between first and second qubits.
The same interaction $\hat{E}=\hat{H}_R\hat{H}_A\hat{CZ}_{A,R}$ is applied between first qubit and the ancilla, and second qubit and the ancilla, respectively.
They are specified with red boxes.
The ancilla is then measured in $\hat{Z}$-basis.
The backaction of this measurement changes the register state $|\psi\rangle$ into
$\hat{X}_1^j\hat{H}_1\hat{H}_2\hat{CZ}_{1,2}|\psi\rangle$
depending on the measurement result $j=0,1$.
} 
\label{CZ}
\end{figure}

For example, it was shown~\cite{ADQC1,ADQC2} that the interaction
\begin{eqnarray*}
\hat{E}\equiv\hat{H}_A\hat{H}_R\hat{CZ}_{A,R}
\end{eqnarray*}
is sufficient for the ancilla-driven universal quantum computation,
where $\hat{H}_A$ is the Hadamard operation on the ancilla,
$\hat{H}_R$ is the Hadamard operation on a register qubit,
and $\hat{CZ}_{A,R}$ is the Controlled-Z (CZ) gate 
\begin{eqnarray*}
\hat{CZ}_{A,R}\equiv
|0\rangle\langle0|_A\otimes\hat{1}_R
+|1\rangle\langle1|_A\otimes\hat{Z}_R
\end{eqnarray*}
between the ancilla
and the register qubit.
Indeed, the single qubit rotation by $u\in{\mathbb R}$
about $z$-axis (plus the Hadamard correction)
\begin{eqnarray*}
\hat{J}(u)\equiv \hat{H} e^{i\frac{u}{2}\hat{Z}}
\end{eqnarray*}
is implemented as is shown in Fig.~\ref{CZ} (a).
The Hadamard correction is canceled by just implementing 
\begin{eqnarray*}
\hat{J}(0)\hat{J}(u)=e^{i\frac{u}{2}\hat{Z}}.
\end{eqnarray*}
The rotation by $u$ about $x$-axis 
is implemented as
\begin{eqnarray*}
\hat{J}(u)\hat{J}(0)=e^{i\frac{u}{2}\hat{X}}.
\end{eqnarray*}
Therefore, according to the Euler decomposition, any single-qubit rotation is possible by using $\hat{J}(u)$.
The CZ gate between two qubits is also implemented by using
the same interaction $E$ as is shown in Fig.~\ref{CZ} (b).
Any single-qubit rotation plus the CZ gate are sufficient for the universal 
quantum computation.

\begin{figure}[htbp]
\begin{center}
\includegraphics[width=0.4\textwidth]{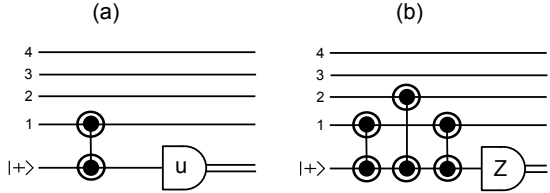}
\end{center}
\caption{Ancilla-driven quantum gates with the CZ+SWAP 
interaction~\cite{ADQC1,ADQC2}.
(a) The quantum circuit realizing the single-qubit rotation $\hat{J}(u)=\hat{H}e^{i\frac{u}{2}\hat{Z}}$.
The CZ+SWAP gate is applied between the qubit we want to rotate 
(say, first qubit)
of the register $|\psi\rangle$ and the ancilla
which is initialized to be $|+\rangle$,
and the ancilla is measured in 
$|u_\pm\rangle\equiv(|0\rangle\pm e^{iu}|1\rangle)/\sqrt{2}$ basis.
The backaction of this measurement changes the register state into
$\hat{X}_1^j\hat{J}_1(u)|\psi\rangle$ depending
on the measurement result $j=0,1$.
(b) The quantum circuit realizing the CZ+SWAP gate between first and second qubits.
The ancilla is measured in $\hat{Z}$-basis.
The backaction of this measurement changes the register state $|\psi\rangle$ into
$(\hat{Z}_1\hat{Z}_2)^j\hat{SWAP}_{1,2}\hat{CZ}_{1,2}|\psi\rangle$
depending on the measurement result $j=0,1$.
} 
\label{CZSWAP}
\end{figure}

In Refs.~\cite{ADQC1,ADQC2}, it was shown that
the interaction
\begin{eqnarray*}
CZ+SWAP\equiv|00\rangle\langle00|+|01\rangle\langle10|
+|10\rangle\langle01|-|11\rangle\langle11|
\end{eqnarray*}
also enables the ancilla-driven universal quantum computation 
(see Fig.~\ref{CZSWAP}).
Interestingly, 
interactions which 
enable the ancilla-driven universal quantum computation
are, apart from local unitaries,
only two types: CZ and CZ+SWAP~\cite{ADQC1,ADQC2}.

\section{Entanglement-fidelity relation for the one-way model}
\label{one-way}

Before giving main results of this paper, let us also review
the entanglement-fidelity relation~\cite{error_cluster}
for the one-way quantum computation,
because, as we will see in the next section,
some of ancilla-driven gates  
have exactly the same entanglement-fidelity relation as
that for the one-way model.

\begin{figure}[htbp]
\begin{center}
\includegraphics[width=0.5\textwidth]{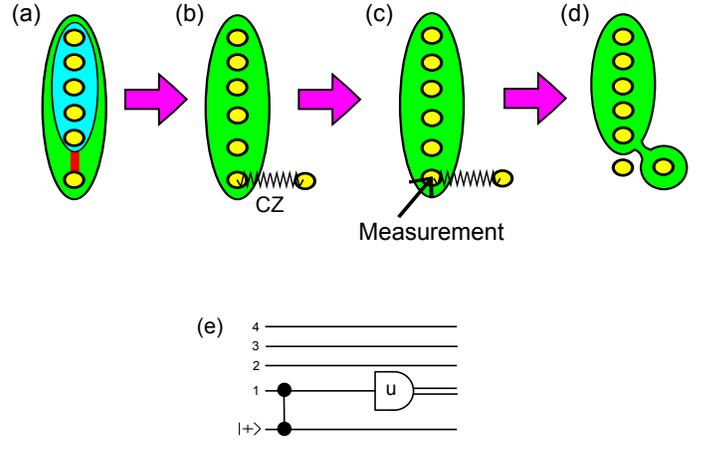}
\end{center}
\caption{(Color online.)
Top line: The elementary process of the one-way single-qubit rotation
considered in Ref.~\cite{error_cluster}.
Yellow circles are qubits. Qubits in the green ellipse play
the role of register qubits.
(a) The initial state. The red bond represents entanglement between the bottommost qubit
and other register qubits (i.e., qubits in the blue ellipse).
(b) The qubit we want to rotate, say the bottommost one,
is coupled to an ancillary qubit through
the CZ interaction (zigzag line).
(c) The bottommost qubit is measured in
$|u_\pm\rangle\equiv(|0\rangle\pm e^{iu}|1\rangle)/\sqrt{2}$ basis (black
solid arrow).
(d) The bottommost qubit is teleported to the ancillary qubit with
the desired single-qubit rotation
$\hat{X}^j\hat{J}(u)$ depending on the measurement result $j=0,1$.
(e) The quantum circuit corresponding to the above elementary process.
} 
\label{single_qubit}
\end{figure}

In Ref.~\cite{error_cluster}, the entanglement-fidelity relation was studied
for the elementary process 
(Fig.~\ref{single_qubit}) of the one-way single-qubit rotation
and the controlled-not gate,
assuming that the projective measurement is inaccurate
in the sense that the direction
to which the qubit is projected is slightly deviated from the
ideal one.
In other words,
the measurement
is not the ideal one $\{|u_+\rangle,|u_-\rangle\}$, where
\begin{eqnarray*}
|u_\pm\rangle\equiv\frac{1}{\sqrt{2}}(|0\rangle\pm e^{iu}|1\rangle),
\end{eqnarray*}
but the slightly deviated one
$\{|\tilde u_+\rangle,|\tilde u_-\rangle\}$, where
\begin{eqnarray*}
|\tilde u_+\rangle&\equiv&
\cos\frac{\epsilon}{2}|u_+\rangle+e^{-i\delta}\sin\frac{\epsilon}{2}|u_-\rangle,\nonumber\\
|\tilde u_-\rangle&\equiv&
\sin\frac{\epsilon}{2}|u_+\rangle-e^{-i\delta}\cos\frac{\epsilon}{2}|u_-\rangle.
\end{eqnarray*}
It is easy to see that the degree of 
the deviation is parametrized by $\epsilon$ and $\delta$:
$|\tilde u_+\rangle$ ($|\tilde u_-\rangle$)
is the vector obtained by rotating $|u_+\rangle$ ($|u_-\rangle$)
by $\epsilon$ about $z$-axis and by $\frac{\pi}{2}-\delta$ 
about $|u_+\rangle$-axis.
If the measurement is accurate (i.e., $\epsilon=\delta=0$), 
the elementary process (Fig.~\ref{single_qubit}) provides 
the single-qubit rotation 
\begin{eqnarray*}
\hat{X}^j\hat{J}(u)
\end{eqnarray*}
with the correction $\hat{X}^j$
depending on the measurement result $j=0,1$.
However, if it is inaccurate ($\epsilon>0,\delta>0$), operation given by
the elementary process is
\begin{eqnarray}
\big(\cos\frac{\epsilon}{2}+(-1)^j\hat{X}
e^{(-1)^ji\delta}\sin\frac{\epsilon}{2}\big)
\hat{X}^j
\hat{J}(u),
\label{error}
\end{eqnarray}
i.e., intuitively, the bit-flip error $\hat{X}$ occurs with the weight
$\sin\frac{\epsilon}{2}$.

Let the register state after the elementary process 
(i.e., the state of qubits in the green ellipse of Fig.~\ref{single_qubit} (d))
be $|\phi_{\epsilon,\delta}\rangle$. If the measurement is accurate,
it is $|\phi_{0,0}\rangle$.
As is shown in Ref.~\cite{error_cluster}, the mean gate fidelity $F$
is explicitly calculated as
\begin{eqnarray}
F\equiv
{\mathbb E}\Big[\big|\langle\phi_{0,0}|\phi_{\epsilon,\delta}\rangle\big|^2\Big]
=\cos^2\frac{\epsilon}{2}+\mbox{Tr}^2(\hat{\rho}_b\hat{Z}_b)\sin^2\frac{\epsilon}{2},
\label{equality}
\end{eqnarray}
where
$\mathbb E[\cdot]$ means the average over all measurement histories,
\begin{eqnarray*}
\hat{\rho}_b\equiv\mbox{Tr}_b(|\psi\rangle\langle\psi|)
\end{eqnarray*}
is the reduced density operator for the bottommost qubit, 
$\mbox{Tr}_b$ is the trace over all register qubits except for the bottommost
qubit, $|\psi\rangle$ is the initial state of the register
(i.e., the state of qubits in the green ellipse of Fig.~\ref{single_qubit} (a)),
and $\hat{Z}_b$ is the Pauli $z$ operator acting on the bottommost qubit.

Let us define the amount $S$ ($0\le S\le 1$) of entanglement
\begin{eqnarray}
S\equiv2[1-\mbox{Tr}(\hat{\rho}_b^2)]
\label{purity}
\end{eqnarray}
between the bottommost qubit
and other register qubits (i.e., qubits in the blue ellipse of Fig.~\ref{single_qubit} (a)).
This entanglement is indicated by the red bond in Fig.~\ref{single_qubit} (a). 
If the bottommost qubit and other register qubits are not entangled,
$S=0$, whereas if they are maximally entangled, $S=1$.

By noticing that $1-S$ is equal to
the square length of the Bloch vector for $\hat{\rho}_b$: 
\begin{eqnarray*}
1-S=
\mbox{Tr}^2(\hat{X}_b\hat{\rho}_b)
+\mbox{Tr}^2(\hat{Y}_b\hat{\rho}_b)
+\mbox{Tr}^2(\hat{Z}_b\hat{\rho}_b),
\end{eqnarray*}
we obtain the obvious inequality
\begin{eqnarray*}
1-S\ge
\mbox{Tr}^2(\hat{Z}_b\hat{\rho}_b).
\end{eqnarray*}
The intuitive meaning of this inequality is that 
if $\hat{\rho}_b$ is more mixed, i.e., the entanglement is larger,
the length of the Bloch vector (and hence
the $z$-component of the Bloch vector) becomes shorter.
From this inequality, 
an upper bound of the right-hand-side of Eq.~(\ref{equality})
is given as
\begin{eqnarray*}
\cos^2\frac{\epsilon}{2}+\mbox{Tr}^2(\hat{\rho}_b\hat{Z}_b)\sin^2\frac{\epsilon}{2}\le
\cos^2\frac{\epsilon}{2}+(1-S)\sin^2\frac{\epsilon}{2},
\end{eqnarray*}
and hence we finally obtain
the desired inequality
\begin{eqnarray}
F\equiv
{\mathbb E}\Big[\big|\langle\phi_{0,0}|\phi_{\epsilon,\delta}\rangle\big|^2\Big]
\le 1-S\sin^2\frac{\epsilon}{2}.
\label{main}
\end{eqnarray}
Note that 
this inequality gives the optimal upper bound for $F$,  
since the equality is always achieved for any given $S$ ($0\le S\le 1$)
by taking 
\begin{eqnarray*}
\hat{\rho}_b=\frac{\sqrt{1-S}}{2}\hat{Z}_b+\frac{1}{2}\hat{1}_b,
\end{eqnarray*}
where $\hat{1}_b$ is the identity operator on the bottommost qubit.

\begin{figure}[htbp]
\begin{center}
\includegraphics[width=0.4\textwidth]{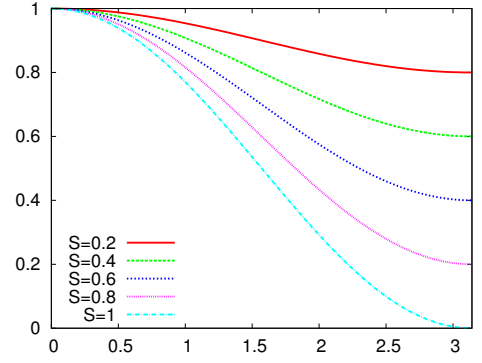}
\end{center}
\caption{
(Color online.)
$1-S\sin^2\frac{\epsilon}{2}$ as a function of $\epsilon$ for
$S=$0.2, 0.4, 0.6, 0.8, and 1.
} 
\label{plot}
\end{figure}

The inequality (\ref{main}) is the relation between entanglement
among register qubits
and the gate fidelity in the inaccurate one-way quantum computation.
The right-hand-side of Eq.~(\ref{main}) is plotted as a function of
$\epsilon$ for various $S$ in Fig.~\ref{plot}.
We can see that for a fixed $\epsilon$, larger $S$ makes $F$ smaller.
As is discussed in Ref.~\cite{error_cluster}, 
$S$ is often very large in many quantum algorithms~\cite{Jozsa,macro_ent}.

Although we have used the purity Eq.~(\ref{purity}) as a measure of entanglement,
we can derive a similar entanglement-fidelity relation by using the von Neumann entropy $S_v$ ($0\le S_v\le 1$)
\begin{eqnarray*}
S_v\equiv-\mbox{Tr}(\hat{\rho}_b\log_2\hat{\rho}_b)
\end{eqnarray*}
as a measure of entanglement. If $\hat{\rho}_b$ is maximally mixed, $S_v=1$,
whereas $S_v=0$ if $\hat{\rho}_b$ is pure.
By a straightforward calculation,
\begin{eqnarray*}
S_v&=&-\frac{1+r}{2}\log_2\frac{1+r}{2}
-\frac{1-r}{2}\log_2\frac{1-r}{2}\\
&\le&-\frac{1+|C_z|}{2}\log_2\frac{1+|C_z|}{2}
-\frac{1-|C_z|}{2}\log_2\frac{1-|C_z|}{2}\\
&\equiv& f(|C_z|),
\end{eqnarray*}
where $r$ is the length 
\begin{eqnarray*}
r\equiv\sqrt{\mbox{Tr}^2(\hat{X}_b\hat{\rho}_b)
+\mbox{Tr}^2(\hat{Y}_b\hat{\rho}_b)
+\mbox{Tr}^2(\hat{Z}_b\hat{\rho}_b)}
\end{eqnarray*}
of the Bloch vector of $\hat{\rho}_b$
and 
\begin{eqnarray*}
C_z\equiv\mbox{Tr}(\hat{\rho}_b\hat{Z}_b).
\end{eqnarray*}
Let $f^{-1}$ be the inverse of $f$. Then,
\begin{eqnarray*}
|C_z|\le f^{-1}(S_v).
\end{eqnarray*}
and therefore
\begin{eqnarray*}
|C_z|^2\le (f^{-1}(S_v))^2.
\end{eqnarray*}
Inserting this inequality into Eq.~(\ref{equality}), we obtain
\begin{eqnarray*}
F \le 1-[1-(f^{-1}(S_v))^2]\sin^2\frac{\epsilon}{2}.
\end{eqnarray*}
The equality in this inequality is always achievable for any $S_v$, 
and therefore this is the optimal upper bound.

\begin{figure}[htbp]
\begin{center}
\includegraphics[width=0.4\textwidth]{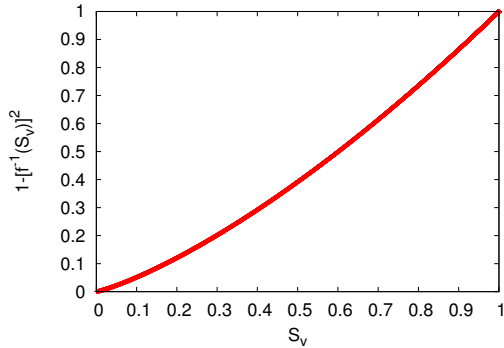}
\end{center}
\caption{
(Color online.)
$1-[f^{-1}(S_v)]^2$ as a function of $S_v$.
} 
\label{inverse}
\end{figure}

In Fig.~\ref{inverse}, we plot
$1-(f^{-1}(S_v))^2$ as a function of $S_v$.
We can see that $1-(f^{-1}(S_v))^2$ is a monotonically increasing function of $S_v$.
Therefore, we can also say that if the entanglement is strong
in terms of $S_v$, the inaccurate measurements
make the gate fidelity low.

\section{Entanglement-fidelity relation for the ancilla-driven model}
\subsection{Single-qubit rotation}
\label{ancilla-drivenrotation}
Let us explore a similar entanglement-fidelity relation
for the ancilla-driven single-qubit rotation
(Fig.~\ref{CZ} (a) and Fig.~\ref{CZSWAP} (a)).
As is mentioned before, we obtain the same inequality Eq.~(\ref{main})
for the ancilla-driven single-qubit rotation. 

In order to see it,
let us first consider the ancilla-driven single-qubit rotation with the
CZ+SWAP interaction (Fig.~\ref{CZSWAP} (a)). 
It is immediate to see that this circuit
is equivalent to that in Fig.~\ref{single_qubit} (e). 
Therefore, we obtain the same inequality, Eq.~(\ref{main}).

Second, let us consider
the ancilla-driven single-qubit rotation with the
CZ interaction (Fig.~\ref{CZ} (a)).
Again, we can show that this circuit is equivalent to
that in Fig.~\ref{single_qubit} (e), since
the application of $\hat{H}_R\hat{H}_A$
on states $|0\rangle_R|+\rangle_A$
or $|1\rangle_R|-\rangle_A$
is equivalent to that of SWAP operation on the same states:
\begin{eqnarray*}
\hat{H}_R\hat{H}_A
|0\rangle_R|+\rangle_A&=&|+\rangle_R|0\rangle_A
=\hat{SWAP}_{R,A}|0\rangle_R|+\rangle_A,\nonumber\\
\hat{H}_R\hat{H}_A
|1\rangle_R|-\rangle_A&=&|-\rangle_R|1\rangle_A
=\hat{SWAP}_{R,A}|1\rangle_R|-\rangle_A.
\end{eqnarray*}

In short, 
Eq.~(\ref{main}) also gives the entanglement-fidelity relation 
for the ancilla-driven single-qubit rotation.
In this case, $S$ is the amount of entanglement between the
qubit we want to rotate and other register qubits in the initial state.

\subsection{Two-qubit entangling gate with the CZ interaction}
\label{ancilla-drivenCZ}

Let us next consider the ancilla-driven two-qubit entangling gate.
We first consider the CZ
interaction (Fig.~\ref{CZ} (b)).
Let the input state  
be
\begin{eqnarray}
|\psi\rangle_r|+\rangle_a=
\Big[\sum_{z_1=0}^1\sum_{z_2=0}^1
\eta_{z_2,z_1}
|\eta_{z_2,z_1}\rangle_o
|z_2\rangle_2|z_1\rangle_1
\Big]|+\rangle_a,
\label{input}
\end{eqnarray}
where $|\psi\rangle_r$ is the register state, 
$|+\rangle_a$ is the
ancilla state,
$\eta_{z_1,z_2}\in{\mathbb C}$,
$|z_i\rangle_i$ is the state of $i$th qubit in the register,
and $|\eta_{z_2,z_1}\rangle_o$ is the state of register qubits
other than first and second qubits.
By a straightforward calculation, the state immediately before the measurement of the ancilla is
\begin{eqnarray*}
&&\eta_{00}|\eta_{00}\rangle_o|+\rangle_2|+\rangle_1|+\rangle_a
+\eta_{01}|\eta_{01}\rangle_o|+\rangle_2|-\rangle_1|-\rangle_a\\
&&+\eta_{10}|\eta_{10}\rangle_o|-\rangle_2|+\rangle_1|+\rangle_a
-\eta_{11}|\eta_{11}\rangle_o|-\rangle_2|-\rangle_1|-\rangle_a.
\end{eqnarray*}
If the measurement is accurate, we obtain the desired output
\begin{eqnarray*}
|\phi_{0,0}^j\rangle\equiv
\hat{X}_1^j\hat{H}_1\hat{H}_2\hat{CZ}_{1,2}|\psi\rangle_r
\end{eqnarray*}
with the correction $\hat{X}_1^j$ which depends on the measurement result $j=0,1$.
However, if the measurement is inaccurate in the sense that
the ancilla is projected onto 
\begin{eqnarray*}
|\tilde{0}\rangle&\equiv&\cos\frac{\epsilon}{2}|0\rangle
+\sin\frac{\epsilon}{2}e^{-i\delta}|1\rangle\\
|\tilde{1}\rangle&\equiv&\sin\frac{\epsilon}{2}|0\rangle
-\cos\frac{\epsilon}{2}e^{-i\delta}|1\rangle,
\end{eqnarray*}
we can show by a straightforward calculation that the output state is
\begin{eqnarray*}
|\phi_{\epsilon,\delta}^j\rangle\equiv
\big(\cos\frac{\epsilon}{2}+(-1)^j\hat{X}_1
e^{(-1)^ji\delta}\sin\frac{\epsilon}{2}\big)
|\phi_{0,0}^j\rangle_r.
\end{eqnarray*}
Note that we obtain the same error operation as that in Eq.~(\ref{error}).
Therefore, by a similar calculation as that for
the derivation of Eq.~(\ref{main}),
we obtain Eq.~(\ref{main}).
In summary, Eq.~(\ref{main}) is also the entanglement-fidelity relation
for the ancilla-driven two-qubit entangling gate with the CZ interaction.

\subsection{Two-qubit entangling gate with the CZ+SWAP interaction}
\label{ancilla-drivenCZSWAP}

Finally, let us consider the CZ+SWAP
interaction (Fig.~\ref{CZSWAP} (b)).
In this case, the entanglement-fidelity relation we will obtain 
is Eq.~(\ref{main2}), which is different from Eq.~(\ref{main})
in the sense that the two-body entanglement appears. However we can still
say from Eq.~(\ref{main2}) that the strong entanglement causes low
gate fidelity.

We assume the same input state Eq.~(\ref{input}).
The state immediately before the final measurement is
straightforwardly calculated as
\begin{eqnarray*}
&&\eta_{00}|\eta_{00}\rangle|0\rangle_2|0\rangle_1|+\rangle_a
+\eta_{01}|\eta_{01}\rangle|1\rangle_2|0\rangle_1|-\rangle_a\\
&&+\eta_{10}|\eta_{10}\rangle|0\rangle_2|1\rangle_1|-\rangle_a
-\eta_{11}|\eta_{11}\rangle|1\rangle_2|1\rangle_1|+\rangle_a.
\end{eqnarray*}
If the measurement is accurate, we obtain
\begin{eqnarray*}
|\phi_{0,0}^j\rangle&\equiv&
(\hat{Z}_1\hat{Z}_2)^j\hat{SWAP}_{1,2}\hat{CZ}_{1,2}|\psi\rangle.
\end{eqnarray*}
If the measurement is inaccurate, we obtain
\begin{eqnarray*}
|\phi_{\epsilon,\delta}^j\rangle&\equiv&
\big(\cos\frac{\epsilon}{2}+
(-1)^j\hat{Z}_1\hat{Z}_2\sin\frac{\epsilon}{2}e^{(-1)^ji\delta}\big)
|\phi_{0,0}^j\rangle.
\end{eqnarray*}
Note that in this case, the error is, intuitively, the
application of $\hat{Z}_1\hat{Z}_2$ with the weight $\sin\frac{\epsilon}{2}$.
This is slightly different from Eq.~(\ref{error}).
By a similar calculation as that for deriving Eq.~(\ref{main}),
we obtain 
\begin{eqnarray*}
F=\cos^2\frac{\epsilon}{2}+\mbox{Tr}^2(\hat{Z}_1\hat{Z}_2\hat{\rho}_{1,2})\sin^2\frac{\epsilon}{2},
\end{eqnarray*}
where 
$\hat{\rho}_{1,2}$
is the reduced density operator for first and second qubits of the
input state of the register.

As a measure of entanglement, let us adopt the von Neumann entropy 
$S_{v2}$ $(0\le S_{v2}\le2)$
\begin{eqnarray*}
S_{v2}(\hat{\rho})=-\mbox{Tr}(\hat{\rho}\log_2\hat{\rho}),
\end{eqnarray*}
where $\hat{\rho}$ is a two-qubit state. The subscript 2 of $S_{v2}$ indicates
two-qubit entanglement. If a two-qubit state $\hat{\rho}$ is maximally
entangled with other qubits, $S_{v2}(\hat{\rho})=2$, 
whereas if it is separable from other qubits, $S_{v2}(\hat{\rho})=0$.

As is shown in Appendix, 
\begin{eqnarray}
S_{v2}(\hat{\rho})&\le&
-\frac{1+|C_{zz}|}{2}\log_2\Big(\frac{1+|C_{zz}|}{4}\Big)\nonumber\\
&&-\frac{1-|C_{zz}|}{2}\log_2\Big(\frac{1-|C_{zz}|}{4}\Big)\nonumber\\
&\equiv& g(|C_{zz}|)
\label{jonas}
\end{eqnarray}
for any $\hat{\rho}$, where
\begin{eqnarray*}
C_{zz}\equiv\mbox{Tr}(\hat{\rho}\hat{Z}\otimes\hat{Z}).
\end{eqnarray*}
Let us denote the inverse of $g$ by $g^{-1}$. Then, 
Eq.~(\ref{jonas}) gives
\begin{eqnarray*}
|C_{zz}|\le g^{-1}(S_{v2})
\end{eqnarray*}
for $1\le S_{v2}\le 2$ (note that $g^{-1}$ is not defined for $0\le S_{v2}\le 1$).
Thus we obtain
\begin{eqnarray}
F\le 1-[1-(g^{-1}(S_{v2}))^2]\sin^2\frac{\epsilon}{2}
\label{main2}
\end{eqnarray}
for $1\le S_{v2}\le2$.
The equality in this inequality is always achievable for any $1\le S_{v2}\le2$,
and therefore this is the optimal upper bound.

\begin{figure}[htbp]
\begin{center}
\includegraphics[width=0.4\textwidth]{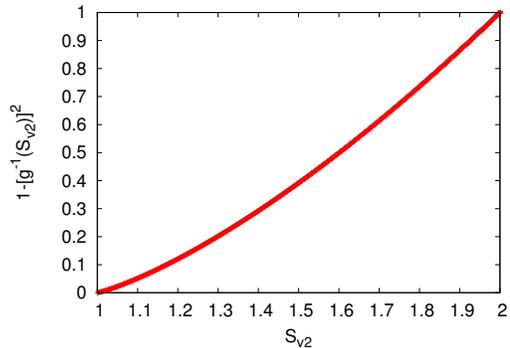}
\end{center}
\caption{
(Color online.)
$1-[g^{-1}(S_{v2})]^2$ as a function of $S_{v2}$.
} 
\label{inverse2}
\end{figure}

In Fig.~\ref{inverse2}, we plot
$1-(g^{-1}(S_{v2}))^2$ as a function of $S_{v2}$. 
We can see that
$1-(g^{-1}(S_{v2}))^2$ is a monotonically increasing function of $S_{v2}$. 
Therefore we can say that if the entanglement is strong 
($S_{v2}\ge1$), the inaccurate measurements
make the gate fidelity low.

Let us also investigate the case $S_{v2}<1$. 
In this case, we cannot say anything about $F$,
since any function of $S_{v2}$ does not provide any nontrivial upper bound
for $|C_{zz}|^2$ if $S_{v2}<1$.
Indeed, let us consider the state
\begin{eqnarray*}
\hat{\rho}_\lambda=\lambda|00\rangle\langle00|+(1-\lambda)|11\rangle\langle11|,
\end{eqnarray*}
where $0\le \lambda \le1$.
Although $S_{v2}(\hat{\rho}_\lambda)$ can take any value between 0 and 1 by changing
$\lambda$, $|C_{zz}|$ is always 1 for any $\lambda$.
Therefore, for $S_{v2}<1$, there is no meaningful 
relation between $S_{v2}$ and $F$.

\section{Summary and discussion}
In this paper, we have studied how the entanglement among register qubits
affects the gate fidelity of the inaccurate ancilla-driven quantum
computation.
By generalizing the previous result about the entanglement-fidelity
relation in the inaccurate one-way quantum computation~\cite{error_cluster},
we have shown similar entanglement-fidelity relations for
the inaccurate ancilla-driven quantum computation.
Our results are
(I) For the ancilla-driven single-qubit rotation, 
and
for the ancilla-driven two-qubit entangling gate with the CZ interaction,
we obtain the entanglement-fidelity relation as given in 
Eq.~(\ref{similar}).
(II) For the ancilla-driven two-qubit entangling gate with the CZ+SWAP interaction,
we obtain the entanglement-fidelity relation Eq.~(\ref{main2}) which 
is slightly different from Eq.~(\ref{similar}).
These relations imply that if the entanglement is strong,
the inaccurate measurements make the gate fidelity low
in the ancilla-driven quantum computation.

As is mentioned in Ref.~\cite{error_cluster},
the error model considered here is a kind of error that can ultimately be
recovered with the quantum error-correcting code.
Therefore, our entanglement-fidelity relations are
of use for studying the stability of a bare quantum computation
in order to obtain valuable feedbacks for the study of general fault-tolerant schemes, to develop the made-to-measure error-correcting codes,
to estimate the threshold value of the error-correction,
and to help experimentalists who want to perform proof-of-principle
experiments with few qubits.

It is very important to study the difference between
the one-way model and the ancilla-driven model
from the view point of our entanglement-fidelity relations.
Although these two models share many similarities,
one interesting difference that has been revealed in this
paper is that not only the single-qubit entanglement
but also the two-qubit entanglement are related to the
gate fidelity in the ancilla-driven case.
Therefore, in addition to the physical constraints
come from the specific experimental setups in the
laboratory,
the amount of the two-qubit entanglement
in the register can also be one criterion for choosing the one-way
model or the ancilla-driven model.
Detailed studies according to this direction would be
a subject of the future study.

\appendix*
\section{Proof of Eq.~(\ref{jonas})}
Let 
\begin{eqnarray*}
H(\hat{\rho}\|\hat{\sigma})\equiv
\mbox{Tr}(\hat{\rho}\log_2\hat{\rho})
-\mbox{Tr}(\hat{\rho}\log_2\hat{\sigma})
\end{eqnarray*}
be the relative entropy between $\hat{\rho}$ and $\hat{\sigma}$.
As is well known~\cite{Nielsen}, a Completely-Positive and Trace-Preserving
(CPTP) map $\mathcal{E}$ 
cannot increase the amount of the relative entropy:
\begin{eqnarray}
H(\mathcal{E}(\hat{\rho})\|\mathcal{E}(\hat{\sigma}))\le 
H(\hat{\rho}\|\hat{\sigma}).
\label{H}
\end{eqnarray}
With the notation $\rho_{ab} = \langle ab \vert \hat{\rho} \vert ab \rangle$ for diagonal elements,
let us consider the CPTP map
\begin{eqnarray*}
\mathcal{E}(\hat{\rho})
&=&\rho_{00}|00\rangle\langle00|
+\rho_{01}|01\rangle\langle01|\\
&&+\rho_{10}|10\rangle\langle10|
+\rho_{11}|11\rangle\langle11|.
\end{eqnarray*}
If we take 
\begin{eqnarray*}
\hat{\sigma}=\frac{1}{4}\hat{1}\otimes\hat{1},
\end{eqnarray*}
where $\hat{1}$ is the single-qubit identity operator, and replace within 
Eq.~(\ref{H}), we obtain:
\begin{eqnarray}
\mbox{Tr}(\hat{\rho}\log_2\hat{\rho})&\ge& 
\rho_{00} \log_2(\rho_{00})
+\rho_{11} \log_2(\rho_{11})\nonumber\\
&&+\rho_{01}\log_2(\rho_{01})+
\rho_{10} \log_2(\rho_{10}).
\label{interm}
\end{eqnarray}

Now since 
\begin{eqnarray*}
C_{zz} & = & 2 (\rho_{00} + \rho_{11}) - 1 \\
       & = & 1 - 2 (\rho_{01} + \rho_{10}),
\end{eqnarray*}
and $(x \mapsto x \log_2 x)$ is a convex function, maximum in the right-hand side of Eq.~(\ref{interm})
for fixed $C_{zz}$ is achieved for
\begin{eqnarray*}
\rho_{00} & =  \rho_{11} = & \frac{1+|C_{zz}|}{4}, \\
\rho_{01} & =  \rho_{01} = & \frac{1-|C_{zz}|}{4}.
\end{eqnarray*}
Substituting in Eq.~(\ref{interm}) yields Eq.~(\ref{jonas}).

















































\acknowledgments
We thank C. Butucea, M. Guta, and  J. Anders for discussion,
and the referee of this manuscript for valuable comments and suggestions.
Both authors acknowledge support by
the French Agence Nationale de la
Recherche (ANR) through the grant StatQuant (JC07 07205763).



\end{document}